\documentclass[aps,prb,showpacs,superscriptaddress,twocolumn]{revtex4}
\usepackage{amsmath}
\usepackage{amsfonts}
\usepackage{graphicx}
\usepackage{psfrag}

\begin{document}

\title{Relativistic electron transport through an oscillating barrier: \\wave packet generation and Fano-type resonances}
\author{L\'{o}r\'{a}nt Zs.~Szab\'{o}}
\author{Mih\'{a}ly G.~Benedict}
\author{Attila Czirj\'{a}k}
\author{P\'{e}ter F\"{o}ldi}
\affiliation{Department of Theoretical Physics, University of Szeged, Tisza Lajos k\"{o}r\'{u}t 84, H-6720 Szeged, Hungary}

\begin{abstract}
Transport properties of massive Dirac particles are investigated through an oscillating barrier. The Floquet quasienergies related to the time-dependent potential appear both in transmission and reflection as sidebands around the incoming electron's energy. We take all relevant sidebands into account and present time averaged transmission and reflection probabilities in a wide energy range. Most qualitative features of scattering on a static barrier -- like Klein paradox -- are still visible, but the transmission probability in the evanescent regime observably increases due to the oscillation of the potential. The strongly inelastic scattering process is shown to lead to multiple Fano-type resonances and temporal trapping of the particles inside the oscillating potential. We also present a detailed study of the time evolution of the wave packets generated in the scattering process. Our results can be relevant for graphene with an induced energy gap.
\end{abstract}

\maketitle

\section{Introduction}
Quantum scattering by time-harmonic potentials is an important and vivid research area. It provides deep understanding of a rich variety of interesting and partly unusual phenomena in strongly driven quantum systems. Photon-assisted tunneling\cite{TG63,BM82} is a remarkable example showing that the presence of an alternating field can lead to strongly inelastic processes. Nonrelativistic quantum mechanical scattering on barriers with oscillating height has been investigated intensively -- mainly in the context of traversal time and photon assisted transport, see e.g.~Refs.~[\onlinecite{BM82,MB04,WZ97}]. It has also been proven that Floquet theory\cite{F883,S65} provides an efficient tool for the investigation of various time-dependent scattering processes.\cite{LR99,MB02}
Fano-type resonances\cite{Fano_original,rmp_82_2257_2010} can appear in this context due to transitions between sideband states and bound states. Developments in the experimental techniques during the last decade enable that these results have the potential of direct applications in the rapidly expanding field of meso- and nanoscale quantum devices.\cite{LU07,KLH05}

Much of the theoretical works published so far studied transport through time-dependent potentials in a nonrelativistic framework. \cite{BM82,MB04,WZ97,LR99,burmeister,burmeister2,MB02} Transport related problems with oscillating spin-orbit interaction have been studied e.g.~in Refs.~[\onlinecite{WC07,RCM08,FBKP09}].
A few recent papers,\cite{pla_376_1159_2012,apl_100_183107_2012,jap_111_103717_2012}
treating massless Dirac particle scattering on time-harmonic
potentials, are inspired by the unique electronic dispersion relation of
graphene.\cite{NG_Science_2004,NG_Nature_2005}
In this single layer of hexagonal carbon
atoms -- as it has been demonstrated experimentally\cite{YK09,HU07,tunable_natn_09} -- the carriers exhibit striking relativistic features like
Zitterbewegung,\cite{jpc_23_143201_2011,CS06} Klein paradox\cite{epjb_83_301_2011,rotelli_pra,rotelli_epjc,S11,klein} and  Klein tunneling.\cite{natphys_2_620_2006,apl_100_163121_2012}

Substrate-induced bandgap in epitaxially grown graphene\cite{nat_mat_2003} opened
the way for its usage as an electronic material. This induced bandgap leads to a finite mass for its charge carriers which obey massive Dirac equation, and the energy dispersion relation is no longer linear in momentum. Based on this, Klein tunneling of massive Dirac fermions through a static barrier,\cite{Setare}
and massive Dirac electron tunneling through a time-periodic potential in single layer graphene\cite{PLAK} were studied.

\begin{figure}[tbh]
\includegraphics[width=8.0cm]{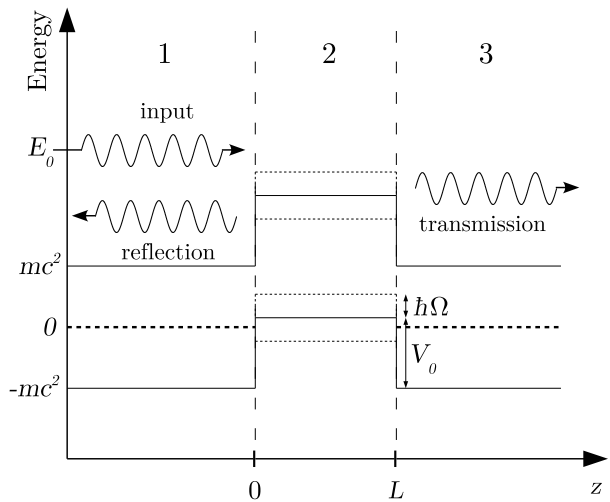}
\caption{Schematic view of the one dimensional scattering problem we consider. The harmonic oscillation of the potential has an amplitude of $\hbar\,\Omega$ and angular frequency of $\nu$. That is,  $V(t)=V_0+\hbar\,\Omega \cos\nu t$ in region 2, while the potential is zero in regions 1 and 3. A monoenergetic electron wave is assumed to impinge the oscillating barrier inducing reflected and transmitted waves in region 1 and 3, respectively. The input energy $E_0$ is always larger than $mc^2$ (half of the induced bandgap), but the magnitudes of $E_0,$ $V_0$ and $\hbar\,\Omega$ relative to each other were varied in our calculations.}
\label{firstfig}
\end{figure}

In the present paper we investigate theoretically the relativistic scattering of massive Dirac particles on a time-periodic rectangular potential barrier (see Fig.\ref{firstfig}),
using Floquet's technique. The method we use is technically similar to that of Ref.~[\onlinecite{PLAK}], but we concentrate on different physical aspects of the problem by presenting spacetime resolved results and discussing Fano-type resonances. Besides its relevance to graphene with a bandgap, this work may also contribute to the manipulation of relativistic charged particle beams by powerful laser pulses.\cite{Nature_431_537_2004,natphys_2_48_2006,natphys_2_696_2006}

Our paper is organized as follows:
in Section II we define the model and derive the equations for the amplitudes, to be solved numerically, from the one-dimensional Dirac equation using Floquet theory. In Section III we  present the physical contents of the numerical solution (e.g. Klein tunneling) in terms of the cycle-averaged and the time-dependent transmission and reflection probabilities. We explore the details of the scattering with the help of space and time dependent charge current and electron density. Finally, we analyze and explain in detail the Fano-type resonances\cite{Fano_original,rmp_82_2257_2010} found in the transmission and reflection curves. We close our paper by summarizing the results and drawing the conclusions in Section IV.

\section{Model}
In this work we consider a two band model of graphene, where the bands are separated by a band gap. This is motivated by the experiments,\cite{nat_mat_2003} where the electronic structure of this unique material has been modified and the degeneracy of the Dirac point at the intersection of the valence and the conduction bands could be removed by growing the sample on a SiC layer. As shown by the measurements, in this case the dispersion relation became similar to that of a massive relativistic Dirac particle, with a band gap $\Delta$ (which appears as $2mc^2$ in the Dirac equation) as large as 0.26\,eV. We consider a one dimensional model where this gap is influenced by a constant plus a harmonically varying potential:
\begin{equation}
V(t)=V_0+V_1(t)=V_0+\hbar\,\Omega \cos\nu t
\end{equation}
which is similar to the gapless model of Ref.~[\onlinecite{jap_111_103717_2012}]. We also note that the nonrelativistic version of a similar problem has been considered in Refs.~[\onlinecite{VE98,Ve97}] in the context of multiphoton ionization. Outside the oscillation region the potential is zero, see Fig.~\ref{firstfig}. A monoenergetic spinpolarized free electron wave is assumed to impinge the oscillating potential barrier, i.e, in the standard representation\cite{greiner} we have
\begin{equation}
\label{in}
 \psi_{\mathrm{in}}(z,t) =e^{ik_0z-i\frac{E_0}{\hbar}t}\left(\begin{array}{c}
                                                   1 \\
                                                   0 \\
                                                   \frac{c\hbar k_0}{E_0 + mc^2} \\
                                                   0
                                                 \end{array}
 \right),
\end{equation}
where the direction of propagation has been chosen to be the $z$-axis. The spinor above is a solution of the Dirac equation with $\hbar k_0=\pm\sqrt{E_0^2 -m^2c^4}/c.$ According to the geometry shown in Fig.~\ref{firstfig}, we choose the positive sign here.

\subsection{Solution of the Dirac equation with oscillating potential}
Inside the region $0<z<L$ the Dirac equation reads
\begin{equation}
\label{tdDirac}
i \hbar \frac{\partial}{\partial t}  \psi (z,t)=H(t)  \psi (z,t),
\end{equation}
with
\begin{equation}
\label{tdHam}
H(t)=H_0+V_1(t)=c\alpha_3\left(-i\hbar \frac{\partial}{\partial z} \right)+ \beta m c^2 + V_0 + V_1(t),
\end{equation}
where the standard $\alpha_3$ and $\beta$ matrices appear. Since $\left[H(t),H(t')\right]=0,$ an eigenstate
\begin{equation}
H_0 \varphi=\mathcal{E} \varphi
\end{equation}
can be used to construct a solution to Eq.~(\ref{tdDirac}):
\begin{equation}
\varphi (t)=\varphi (0)\ e^{-\frac{i}{\hbar}\left(\mathcal{E} t + \int\limits_0^t V_1(\tau)d\tau\right)}.
\end{equation}
For a fixed value of $k,$ (i.e., spatial dependence of $e^{ikz}$), we have
\begin{equation}
\label{Epm}
\mathcal{E}^{\pm}(k) = \pm \sqrt{m^2 c^4 + \hbar ^2 k^2 c^2}+V_0,
\end{equation}
which are both doubly degenerate (due to the two possible spin directions).
Since the interaction is independent of spin [the terms $V_0+V_1(t)$ are proportional to the unit matrix in Eq.~(\ref{tdDirac})], the solutions of Eq.~(\ref{tdDirac}) that correspond to the monoenergetic incoming spinor (\ref{in}) as a boundary condition, have nonzero components at the same positions as $ \psi _{\mathrm{in}}$. Therefore it is sufficient to consider only
\begin{equation}
\label{phipm}
\varphi^\pm(z,t)=e^{i k z}u^\pm(k) \ e^{-{i}(\frac{\mathcal{E}^\pm t}{\hbar}+\frac{\Omega}{\nu}\sin\nu t)},
\end{equation}
\begin{equation}
\label{up}
 u^+ (k) =\left(
    \begin{array}{c}
                                                   1 \\
                                                   0 \\
                                                   \frac{c \hbar k}{\mathcal{E}^+(k) -V_0 + mc^2} \\
                                                   0
                                                 \end{array}
 \right),
 \end{equation}

 \begin{equation}
\label{um}
  u^- (k) =\left(
    \begin{array}{c}
                                                    \frac{c \hbar k}{\mathcal{E}^-(k) -V_0 - mc^2} \\
                                                   0 \\
                                                   1 \\
                                                   0
                                                 \end{array}
 \right)
\end{equation}
that clearly satisfy Eq.~(\ref{tdDirac}). Using Eq.~(\ref{Epm}), it is readily seen that the spinors above do not depend on $V_0.$ Thus e.g.~Eq.~(\ref{in}) we can be rewritten as $ \psi _{\mathrm{in}}(z,t) =e^{ik_0z-i\frac{E_0}{\hbar}t}  u^+ (k_0).$ [In this case $\mathcal{E}^{+}(k_0)=E_0.$]

Note that the only restriction in Eqs.~(\ref{Epm}-\ref{um}) concerning $k$ is that $\hbar^2k^2\geq-m^2c^2$ [ensuring that $\mathcal{E}^{\pm}$ are real, see Eq.~(\ref{Epm})]. This means that evanescent solutions with purely imaginary $k$ are also allowed.

Using the Jacobi-Anger identity
\begin{equation}
\label{JacAng}
e^{-i \frac{\Omega}{\nu}\sin \nu t}=\sum\limits_{n=-\infty}^{\infty}J_n\Bigl(\frac{\Omega}{\nu}\Bigr)\ e^{-i n \nu t},
\end{equation}
where $J_n$ denote Bessel functions of the first kind,
we see that the frequencies appearing in the time evolution are given by
$\mathcal{E}^\pm(k)/\hbar+ n\nu,$
with integer $n$. Note that since the differential operator given by Eq.~(\ref{tdHam}) is periodic in time $(\mathcal{T}=2\pi/\nu),$ Floquet theory\cite{F883,S65} can be applied. The states (\ref{phipm}) are orthogonal in the spinor sense, they can be considered as elements of a time-dependent basis.  Apart from the factors $\exp(-i\mathcal{E}^\pm(k)t/\hbar),$ these solutions are periodic, thus $\mathcal{E}^\pm(k)$ can be called the (nonequivalent) Floquet quasienergies. The term nonequivalent means here that e.g. $\tilde{\mathcal{E}}^-(k)=\mathcal{E}^-(k)+n\hbar\nu$ can also play the role of a Floquet quasienergy (with $n$ being an integer), but states with time-dependences $\exp[-i\mathcal{E}^-(k)t/\hbar-i(\Omega/\nu)\sin(\nu t)]$ and $\exp[-i\tilde{\mathcal{E}}^-(k)t/\hbar-i(\Omega/\nu)\sin(\nu t)+in\hbar\nu]$ are dynamically equivalent. On the other hand, $\mathcal{E}^+(k)$ and $\mathcal{E}^-(k)$ correspond to qualitatively different dynamical behavior (unless their difference is an integer multiple of $\hbar\nu$).

\subsection{Fitting the solutions}
In the previous subsection plane wave solutions of the Dirac equation were obtained. It was shown that for a given (real or purely imaginary) value of $k,$ Eq.~(\ref{tdDirac}) is satisfied by the spinors $\varphi^\pm(z,t)$ given by Eq.~(\ref{phipm}). However, as we shall see, in order to obtain a solution to the problem over the whole $z$-axis, several (in principle infinite number of) different wave vectors are needed.

According to the previous subsection, in region 2 (where the potential oscillates) whenever a frequency $\mathcal{E}/\hbar$ appears in the time evolution, the  harmonics $\mathcal{E}/\hbar+n\nu$ are also present ($n=\ldots, -2, -1, 0, 1, 2, \ldots$). However, if we want to impose continuity of the spinor valued wave function at the boundaries $z=0$ and $z=L,$ the linear equations have nontrivial solutions only if the input frequency equals to one of the  harmonics mentioned above. In other words, the frequencies we have to take into account are
\begin{equation}
\label{omegan}
\omega_n=E_n/\hbar=E_0/\hbar+ n\nu,
\end{equation}
where $E_0$ is the well defined energy of the input spinor, see Eq.~(\ref{in}). In region 1, the only right propagating spinor (see Fig.~\ref{firstfig}) is the input; a particular solution of the Dirac equation corresponding to frequencies $\omega_n$ is given by:
\begin{align}
\Psi_1(z,t)&= \psi _{\mathrm{in}}(z,t) \nonumber\\
&+\sum\limits_{\omega_n>0} r_n\ e^{-i k_n z}u^+(-k_n)\ e^{-i \omega_n t}\nonumber \\
&+\sum\limits_{\omega_n<0} r_n\ e^{-i k_n z}u^-(-k_n)\ e^{-i \omega_n t},
\label{region1}
\end{align}
where
\begin{equation}
k_n=\begin{cases}\sqrt{\frac{E_n^2-m^2c^4}{\hbar^2c^2}},& \mbox{if } E_n^2>m^2c^4 \\ i\sqrt{\frac{m^2c^4-E_n^2}{\hbar^2c^2}},& \mbox{if } E_n^2<m^2c^4.\end{cases}
\end{equation}
The signs here have been chosen such that the terms proportional to $r_n$ in Eq.~(\ref{region1}) describe either reflected or evanescent waves (with exponentially decaying amplitude as $z\rightarrow-\infty.$)
Analogously, in region 3:
\begin{align}
\Psi_3(z,t)&=\sum\limits_{\omega_n>0} t_n\ e^{i k_n z}u^+(k_n)\ e^{-i \omega_n t}\nonumber \\
&+\sum\limits_{\omega_n<0} t_n\ e^{i k_n z}u^-(k_n)\ e^{-i \omega_n t}.
\label{region3}
\end{align}

The problem is somewhat more complicated in region 2, where
\begin{equation}
k'_n=\begin{cases}\sqrt{\frac{(E_n-V_0)^2-m^2c^4}{\hbar^2c^2}},& \mbox{if } (E_n-V_0)^2>m^2c^4 \\ i\sqrt{\frac{m^2c^4-(E_n-V_0)^2}{\hbar^2c^2}},& \mbox{if } (E_n-V_0)^2<m^2c^4 \end{cases}
\label{kleinpar}
\end{equation}
are the solutions of
\begin{equation}
\mathcal{E}^{\pm}(k_n) = E_n.
\end{equation}
However, in this case, due to the oscillation of the potential, a given wave number $k'_n$ corresponds to not only a single $\omega_n.$ For the sake of simplicity, we collect the coefficients of the two types of spinors given by Eqs.~(\ref{up},\ref{um}) separately, and write
\begin{equation}
\Psi_2(z,t)=\Psi_2^+(z,t)+\Psi_2^-(z,t).
\label{region2}
\end{equation}
The first term here is given by
\begin{align}
\Psi_2^+(z,t)&=\sum\limits_{\omega_n>0}  e^{-i \omega_n t -i\frac{\Omega}{\nu}\sin\nu t}  \nonumber \\
&\times \left[ a_n\ e^{i k'_n z}u^+(k'_n) +b_n\ e^{-i k'_n z}u^+(-k'_n)\right],
\end{align}
where, as we can see, both propagation directions appear. Additionally, due to the finite size of the region, exponentially growing spatial dependence is also allowed. Next we insert the Jacobi-Anger expansion (\ref{JacAng}) and obtain an equation where the frequencies $\omega_n$ appear explicitly. For the sake of brevity, we present this step only for the terms containing the spinor $u^-$:
\begin{align}
\Psi_2^-(z,t)&=\sum\limits_{\omega_n<0} \sum_m J_m\left(\frac{\Omega}{\nu}\right)e^{-i \omega_{n+m} t} \nonumber \\
&\times \left[ a_n\ e^{i k'_n z}u^-(k'_n) +b_n\ e^{-i k'_n z}u^-(-k'_n)\right].
\end{align}

The condition of continuity at $z=0$ ($z=L$) now can be formulated by evaluating $\Psi_1(0,t)$ and $\Psi_2(0,t)$ [$\Psi_2(L,t)$ and $\Psi_3(L,t)$]. Working in frequency domain, as an example, the contribution of $\Psi_2^-(z=0)$ to the frequency component $\omega_l$ is given by:
\begin{equation}
\sum\limits_{\omega_n<0} J_{l-n}\left(\frac{\Omega}{\nu}\right)  \left[ a_n u^-(k'_n) +b_n u^-(-k'_n)\right].
\end{equation}
By comparing the coefficients of each $\omega_l$ at the boundaries, we obtain an infinite system of linear equations for the unknown coefficients $\{r_n, t_n, a_n, b_n \}.$ However, since the Bessel functions $J$ for a given argument generally decrease as a function of their index, correct numerical solutions could be obtained by taking only a finite number of frequencies into account. If we consider a set $\{\omega_n: n=-N,\ldots,0,\ldots N\},$ there will be $8N+4$ fitting equations and the same number of unknowns. As we shall see in the next section, the time averages of the transmission and reflection probabilities provide an efficient tool for monitoring the accuracy of the numerical method: if their sum is not as close to unity as we require, $N$ has to be increased.

\section{Results}
The time-dependent reflection and transmission probabilities are given by the ratio of the transmitted and reflected currents to the incoming one:
\begin{align}
R(t)=&\frac{j_{r}(t)}{j_{\mathrm{in}}}=\frac{E_0+mc^2}{2c \hbar k_0} \ \widetilde{\Psi_1^\dagger}\alpha_3\widetilde{\Psi_1}(0,t), \nonumber
\\
T(t)=&\frac{j_{t}(t)}{j_{\mathrm{in}}}=\frac{E_0+mc^2}{2c \hbar k_0} \ \Psi_3^\dagger\alpha_3\Psi_3(L,t),
\end{align}
where
\begin{equation}
\widetilde{\Psi_1}={\Psi_1}- \psi _{\mathrm{in}},
\end{equation}
i.e., the reflected (or exponentially decaying) part of ${\Psi_1}.$ Let us note that the parameter ranges we use in the following are ideal to see and identify the physical processes that are responsible for the effects to be presented. To this end we use "natural units" (i.e., $\hbar=1, m=1$ and $c=1$) for the figures.

\subsection{Cycle-averaged reflection and transmission probabilities}

\begin{figure}[tbh]
\includegraphics[width=8.0cm]{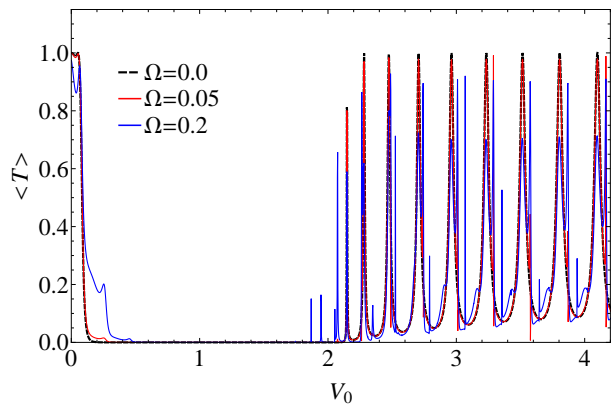}
\caption{$\left\langle T \right\rangle$ as a function of barrier height $V_0$ for the indicated values of the oscillation amplitude $\Omega$. Additional parameters: $E_0=1.1 mc^2, \nu=0.2, L=10$. The physical meaning of the spikes will be discussed later in Sec.~III C.}
\label{fig2}
\end{figure}

Since the time dependence of both $T$ and $R$ contains factors $\exp[-i(\omega_n-\omega_m)t]=\exp[-i\nu(n-m)t]$, these functions are periodic, $T(t+\mathcal{T})=T(t),$  $R(t+\mathcal{T})=R(t),$ with $\mathcal{T}=2\pi/\nu.$ First we consider the time average of the reflection and transmission probabilities,
\begin{equation}
\left\langle T\right\rangle=\int_0^{\mathcal{T}}T(t) dt,\ \ \  \left\langle R \right\rangle=\int_0^{\mathcal{T}}R(t) dt.
\end{equation}
Using Eqs.~(\ref{region1}) and (\ref{region3}), we have
\begin{align}
\left\langle T \right\rangle&=\sum_{\Im(\omega_n)=0} \left|t_n \right|^2 \frac{2c \hbar k_n}{E_n+mc^2},\nonumber \\
\left\langle R \right\rangle&=\sum_{\Im(\omega_n)=0} \left|r_n \right|^2 \frac{2c \hbar k_n}{E_n+mc^2}.
\end{align}
Note that since the scattering problem is periodic in time, the cycle-average of the incoming current should be equal to
$\left\langle j_t \right\rangle-\left\langle j_r\right\rangle=\left|\left\langle j_t\right\rangle\right|+\left|\left\langle j_r\right\rangle\right|.$ In other words,
\begin{equation}
\label{rplust}
\left\langle T \right\rangle +\left\langle R \right\rangle=1
\end{equation}
should hold. This requirement can be used to monitor the accuracy of our calculations. Since we truncate the infinite system of equations, it is not necessary, that Eq.~(\ref{rplust}) is satisfied. However, if the populations of the states that are neglected due to the truncation are negligible, the error can be kept below an acceptable limit. For the results to be presented in the following,  $\left|1-\left\langle T \right\rangle-\left\langle R\right\rangle\right|\leq 10^{-6}$, and to achieve this limit it was sufficient to truncate the system at $N=20-25$ depending on various parameters.
\begin{figure}[tbh]
\includegraphics[width=8.0cm]{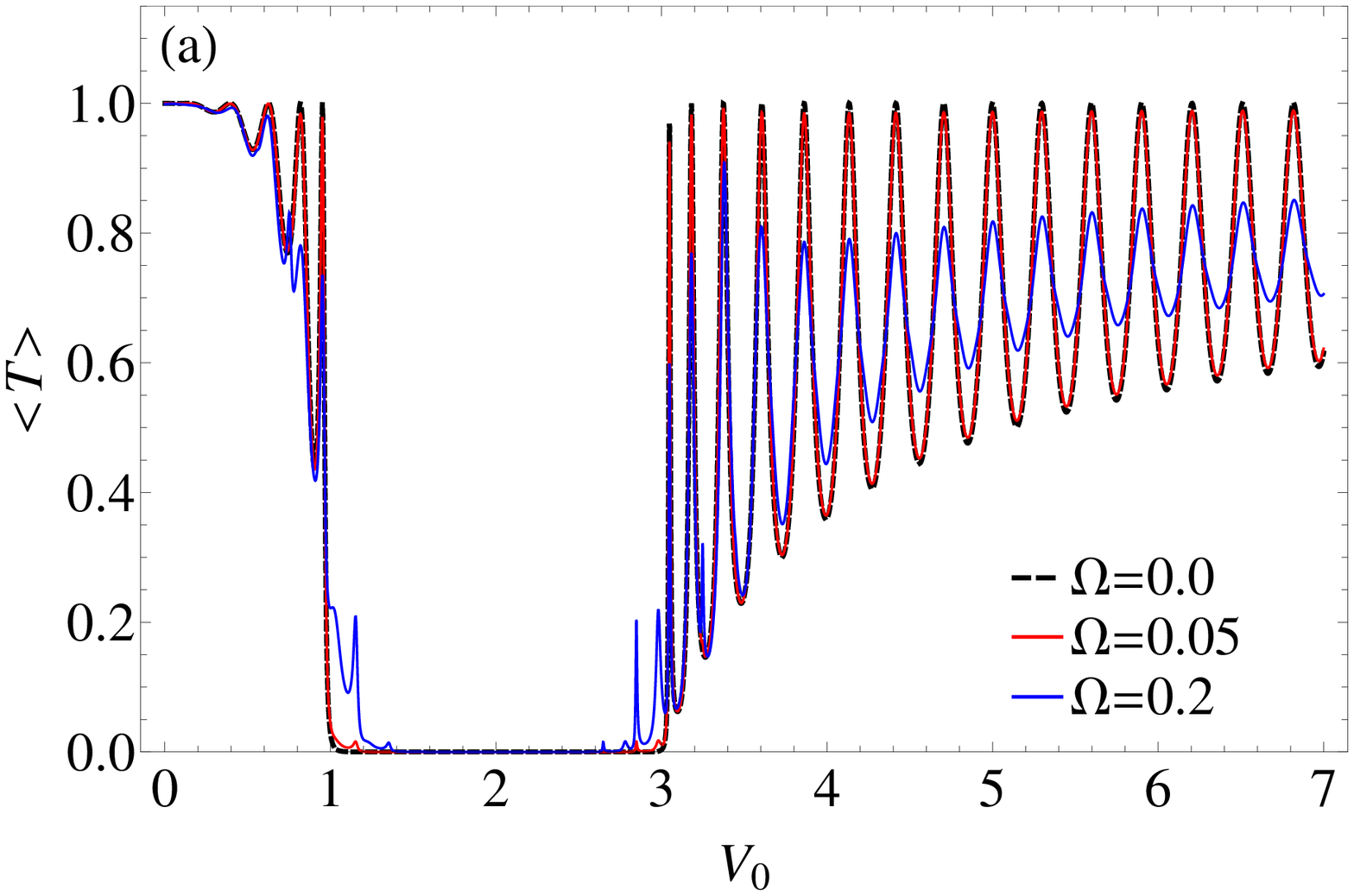}
\includegraphics[width=8.0cm]{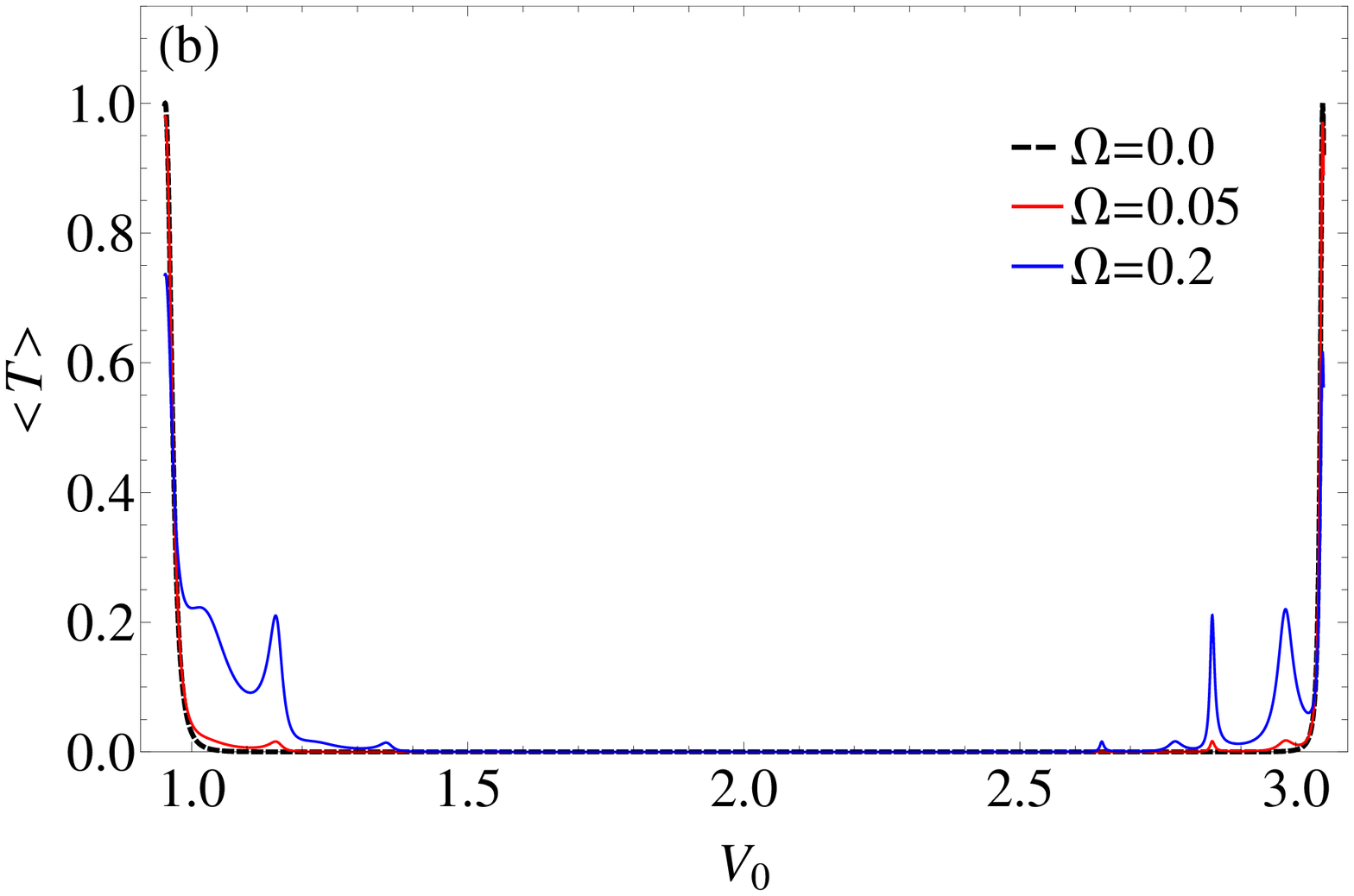}
\caption{Panel (a): $\left\langle T \right\rangle$ as a function of barrier height $V_0$ for the indicated values of the oscillation amplitude $\Omega$. Additional parameters: $E_0=2 mc^2, \nu=0.2, L=10$.
Panel (b): Zoom into the interval $E_0-mc^2<V_0<E_0+mc^2$ of panel (a).}
\label{fig3}
\end{figure}

\begin{figure}[tbh]
\includegraphics[width=8.0cm]{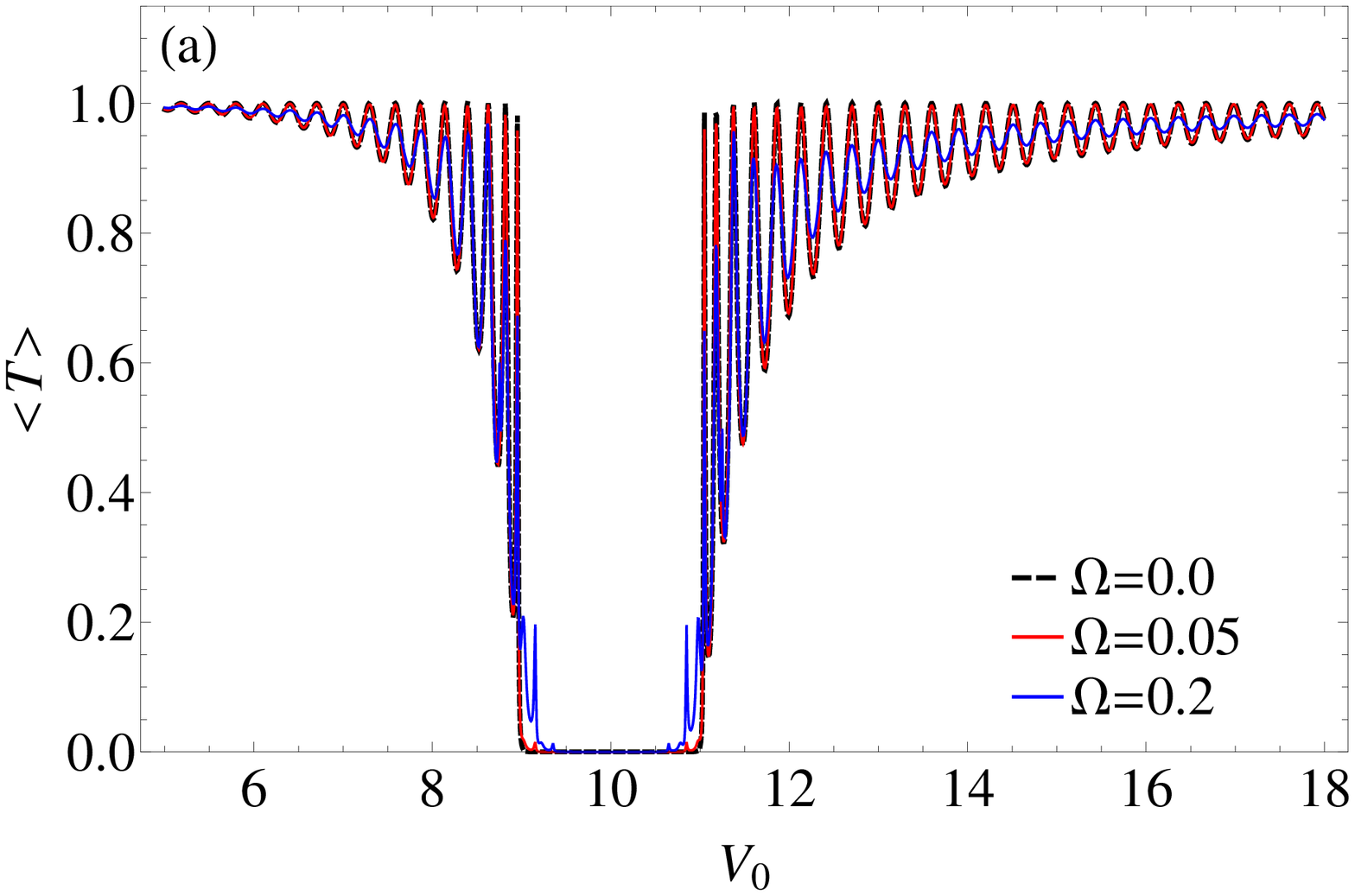}
\includegraphics[width=8.0cm]{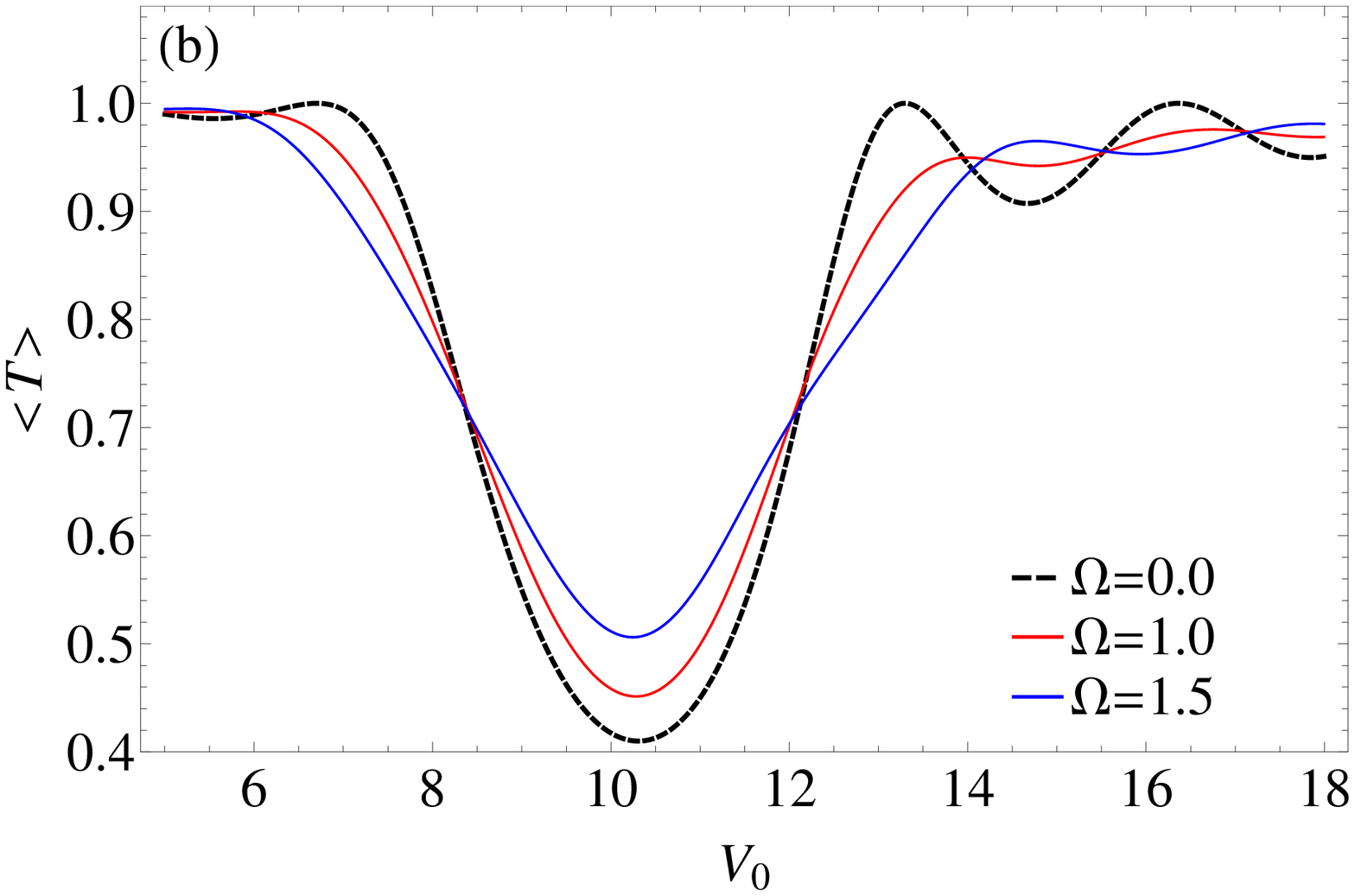}
\caption{$\left\langle T \right\rangle$ as a function of barrier height $V_0$ for the indicated values of the oscillation amplitude $\Omega$.
Additional parameters in panel (a): $E_0=10 mc^2, \nu=0.2, L=10$,
panel (b): $E_0=10 mc^2, \nu=0.2, L=1$.}
\label{fig4}
\end{figure}

Figs.~\ref{fig2}-\ref{fig4} show $\left\langle T \right\rangle$ as a function of $V_0$ for the weakly relativistic, relativistic and ultrarelativistic cases (when $E_0$ is close to $mc^2$, $E_0=2mc^2$ and $E_0=10mc^2$, respectively).  The dashed black curve -- as a reference -- corresponds to the case of $\Omega=0$ (non-oscillating barrier) in all figures. The most important point that Figs.~\ref{fig2}-\ref{fig4} have in common is that for $\Omega=0$ (and oscillations with small amplitude) $\left\langle T\right\rangle$ is practically unity when $V_0\ll E_0;$ it is almost zero when $E_0-mc^2<V_0<E_0+mc^2$ and converges (in an oscillating way) to 1 again, when $V_0>E_0+mc^2.$ This is a well known behavior, that can be understood readily by investigating Eqs.~(\ref{kleinpar}): The solutions in region 2 (see Fig.~\ref{firstfig}) are propagating waves in the first case, real exponentials in the evanescent domain ($E_0-mc^2<V_0<E_0+mc^2$) and propagating waves again when $V_0>E_0+mc^2.$ In other words, $\left\langle T\right\rangle$ reproduces the Klein paradox\cite{epjb_83_301_2011,rotelli_pra,rotelli_epjc,S11,klein} for small values of $\Omega.$

The oscillations that can be seen in Figs.~\ref{fig2}-\ref{fig4} are signatures of quantum mechanical interference: there are maxima (minima) in $\left\langle T\right\rangle$ when the spinor valued waves interfere constructively (destructively) at $z=L.$ When the ratio $\Omega/\nu$ is small, terms corresponding to wavenumbers $k_0$ and $k'_0$ dominate the dynamics. For larger values of $\Omega/\nu,$ the expansion (\ref{JacAng}) in terms of Bessel functions contains numerous frequency components resulting in more complex oscillation patterns in Figs.~\ref{fig2}-\ref{fig4}.
\bigskip

The figures show an additional, important effect, namely the gradual disappearance of the pronounced flat minimum of $\left\langle T\right\rangle(V_0)$ as either $L$ is decreased, or $\Omega$ increased. The first case is related to the role of the evanescent solutions, since tunneling becomes increasingly efficient when the width of region 2 is decreased. When the amplitude of the potential oscillations is increased, more and more frequency components play a relevant role in the dynamics. Some of them corresponds to (quasi)energies $E_n$ that are higher than the oscillating barrier, and consequently the related part of the spinor valued waves are transmitted with a high probability.

\begin{figure}[tbh]
\includegraphics[width=8.0cm]{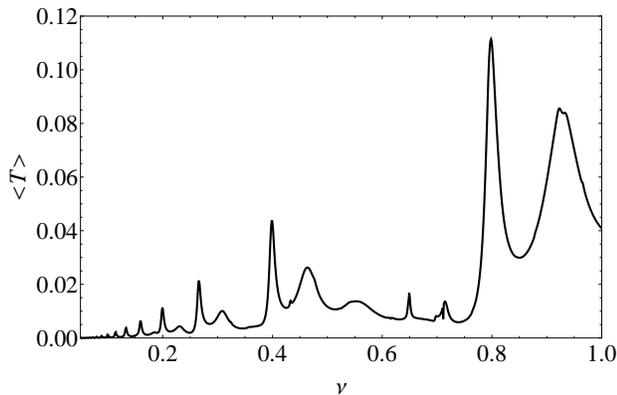}
\caption{$\left\langle T \right\rangle$ as a function of $\nu$,
for parameters $E_0=3 mc^2, V_0=2.75, L=10, \Omega=0.3$.
Note that the transmission probability in the static case for the same parameters
(evanescent solutions) has the order of magnitude of $10^{-8}$. Note that the physical relevance of the abrupt changes seen in this figure will be analyzed in detail in subsection \ref{Fanosubsec}.}
\label{fig5}
\end{figure}

This effect is still present in the case when the energy of the incoming spinor, $E_0,$ is below the minimum of the oscillating potential:
\begin{equation}
E_0-mc^2<V_0-\hbar\,\Omega.
\label{lowE0}
\end{equation}
Without oscillations, $\left\langle T\right\rangle=T$ would be practically zero in this case.  Fig.~\ref{fig5} corresponds to parameter values where Eq.~(\ref{lowE0}) is satisfied and $T<10^{-7}$ when $\Omega=0.$ However, as we can see in the figure, orders of magnitude higher cycle-averaged transmission probabilities arise when $\Omega=0.3$ (in natural units.) The dependence of $\left\langle T\right\rangle$ on the frequency $\nu$ shown in Fig.~\ref{fig5} tells us that although lower values of $\nu$ with a fixed $\Omega$ means that more frequencies $\omega_n$ should be taken into account, this effect is weaker than the fact that higher values of $\nu$ corresponds to larger steps in the ladder $\omega_n=\omega_0+n\nu.$ That is, as a tendency, $\left\langle T\right\rangle(\nu)$ is an increasing function in the parameter range given by Eq.~(\ref{lowE0}).

\subsection{Spacetime dependence of the process: wave packet generation and propagation}
It is instructive to investigate the space and time dependent quantities
\begin{equation}
\label{rhoj}
\rho(z,t)=\Psi^\dagger(z,t)\Psi(z,t), \  \  j(z,t)=c \Psi^\dagger(z,t)\alpha_3\Psi(z,t)
\end{equation}
that satisfy the continuity equation
\begin{equation}
\label{cont}
\frac{\partial}{\partial t}\rho(z,t)=-\frac{\partial}{\partial z}j(z,t).
\end{equation}
The spinor $\Psi$ in Eq.~(\ref{rhoj}) above stands for $\Psi_i, \ i=1,2,3$, depending on whether $z$ is in region 1, 2 or 3, respectively. Note that for $\Omega=0,$ $\rho$ does not depend on time, and consequently $j$ is constant as a function of $z.$ Considering an oscillating potential, the spacetime dependence of both quantities are considerably more interesting, and sheds light on the cycle-averaged results presented in the previous subsection.
\begin{figure}[tbh]
\includegraphics[width=8.0cm]{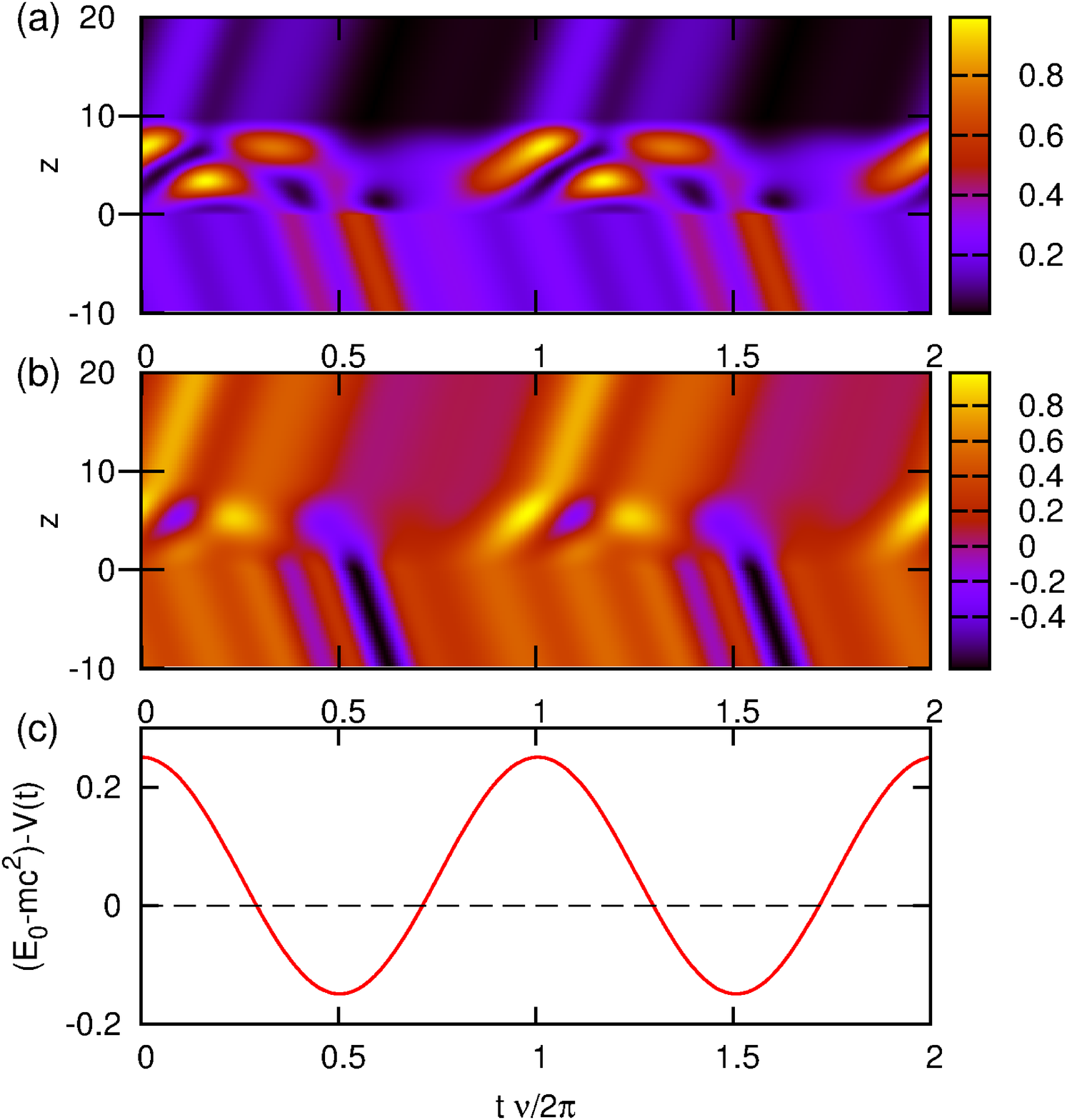}
\caption{Density plot of $\rho(z,t)$ in panel (a), and of $j(z,t)$ in panel (b), and plot of $E_0-mc^2-V(t)$ in panel (c), for the parameters $L=10$, $V_0=1.95 mc^2$, $E_0=3 mc^2$, and $\Omega=0.2$.}
\label{fig6}
\end{figure}

\begin{figure}[tbh]
\includegraphics[width=8.0cm]{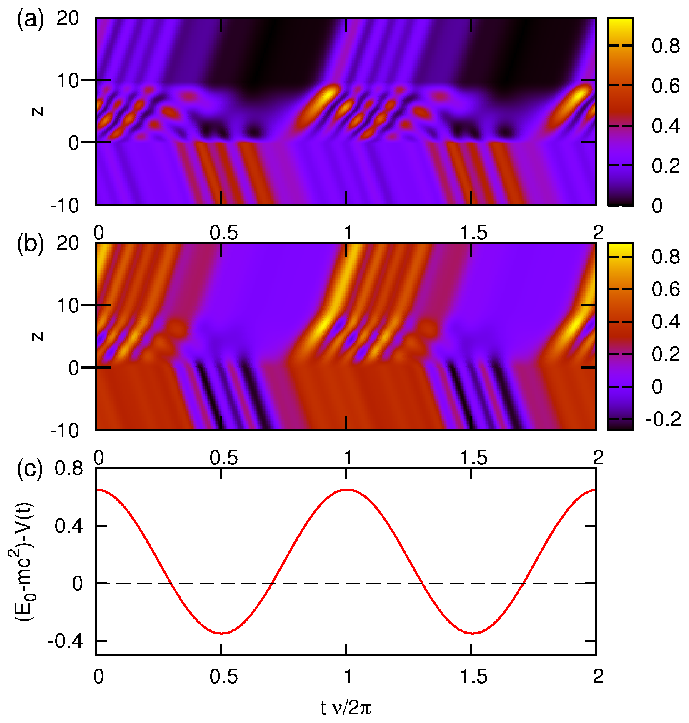}
\caption{Density plot of $\rho(z,t)$ in panel (a), and of $j(z,t)$ in panel (b), and plot of $E_0-mc^2-V(t)$ in panel (c), for the parameters $L=10$, $V_0=1.85 mc^2$, $E_0=3 mc^2$, and $\Omega=0.5$.}
\label{fig7}
\end{figure}

The physical phenomena being responsible for the results presented so far are most visible by focusing on parameters yielding $\left\langle T\right\rangle=\left\langle R\right\rangle=1/2.$ Figs.~\ref{fig6} and \ref{fig7} show $\rho(z,t)$ and $j(z,t)$ in this case, for different values of $\Omega.$ As previously, the patterns that can be seen are more complex for larger amplitude of the potential oscillations, but the qualitative features of Figs.~\ref{fig6} and \ref{fig7} are the same. Considering the electron density, it has certain pronounced maxima in region 2 (the value of $\rho$ at these points can be larger than anywhere else by a factor of two), i.e., we observe a kind of temporal "trapping" of the population inside the oscillating barrier.

We can also see structured wave packets that propagate in the positive (negative) $z$ direction in region 3 (1) due to the "pumping" (see e.g.~Refs.~[\onlinecite{MB02,RCM08,PC13,FA13}]) caused by the oscillating barrier. Considering the time intervals when these wave packets are released, it is instructive to see panel c) in Figs.~\ref{fig6} and \ref{fig7}, where $E_0-mc^2-V(t)$ is plotted as a function of time. When this quantity is positive, the potential barrier is lower than $E_0,$ and we see wave packets leaving the central region in the forward direction (transmission). Oppositely, when $E_0-mc^2-V(t)<0,$ we have mainly reflection.

The current density's spacetime behavior resembles that of $\rho,$ but in this case the sign contains an additional information. In region 3, $j$ is by construction always positive, although its magnitude varies. In this region $\rho$ and $j$ have very similar spacetime dependence. For $z<0,$ however, both the incoming and the reflected spinors contribute to the current density, and their interference can result in negative or positive $j$ (depending on whether $\widetilde{\Psi_1}$ or $\psi_{\mathrm{in}}$ is the dominant, respectively.) Note that when $\left\langle R\right\rangle\approx 1,$ $j(z,t)$ is practically zero except region 1, where it represents a truly alternating current, with $\max[j(t)]\approx-\min[j(t)].$ Obviously, in this case the cycle average of the current is zero everywhere.

\subsection{Fano-type resonances}
\label{Fanosubsec}
As it is known, whenever the energy of a scattering state coincides with that of a bound state, or in other words, these two different kinds of eigenstates happen to belong to the same degenerate energy level, then the transition probability between the states becomes large and a resonance occurs in the transmission. In general, this effect is known as a Fano resonance.\cite{Fano_original,rmp_82_2257_2010}

The role of Fano-type resonances has been discussed earlier in the context of the model of gapless graphene,\cite{jap_111_103717_2012} which has a different dispersion relation than the one discussed here.
Ref.~[\onlinecite{jap_111_103717_2012}] presents a detailed analysis of these resonances for the case of a
potential well, in the context of a two-dimensional massless Dirac equation.
The way the plots of the transmission probability vs. incident energy in Ref.~[\onlinecite{jap_111_103717_2012}] depend on the most important model parameters is a description of the Fano-type resonances which is complementary to ours.
By calculating the Wigner delay time,\cite{WIG55,SM60} they also predict the temporal trapping that we are going to show below explicitly.

In our present case a Fano-type resonance may occur if the energy of the scattered electron,
after losing a number of quanta $n\hbar\nu$,
 coincides with the bound state energy $\epsilon$ of a Dirac particle in the time-independent potential,
 lying between $-mc^2$ and $mc^2$:
\begin{equation}
E_0+n\hbar\nu=\epsilon, \ n=-1, -2,...
\end{equation}
\begin{figure}[tbh]
\includegraphics[width=8.0cm]{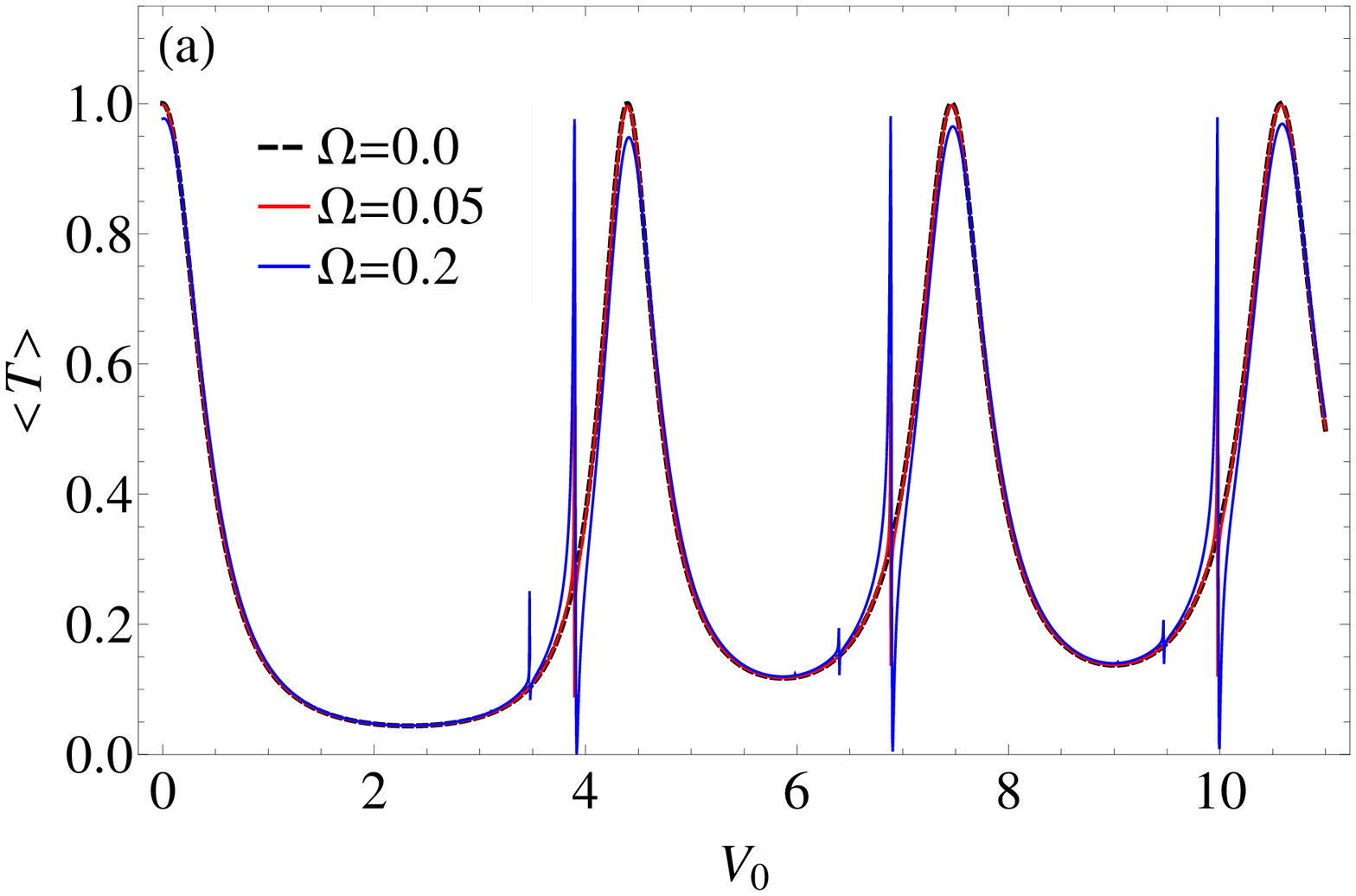}
\includegraphics[width=8.0cm]{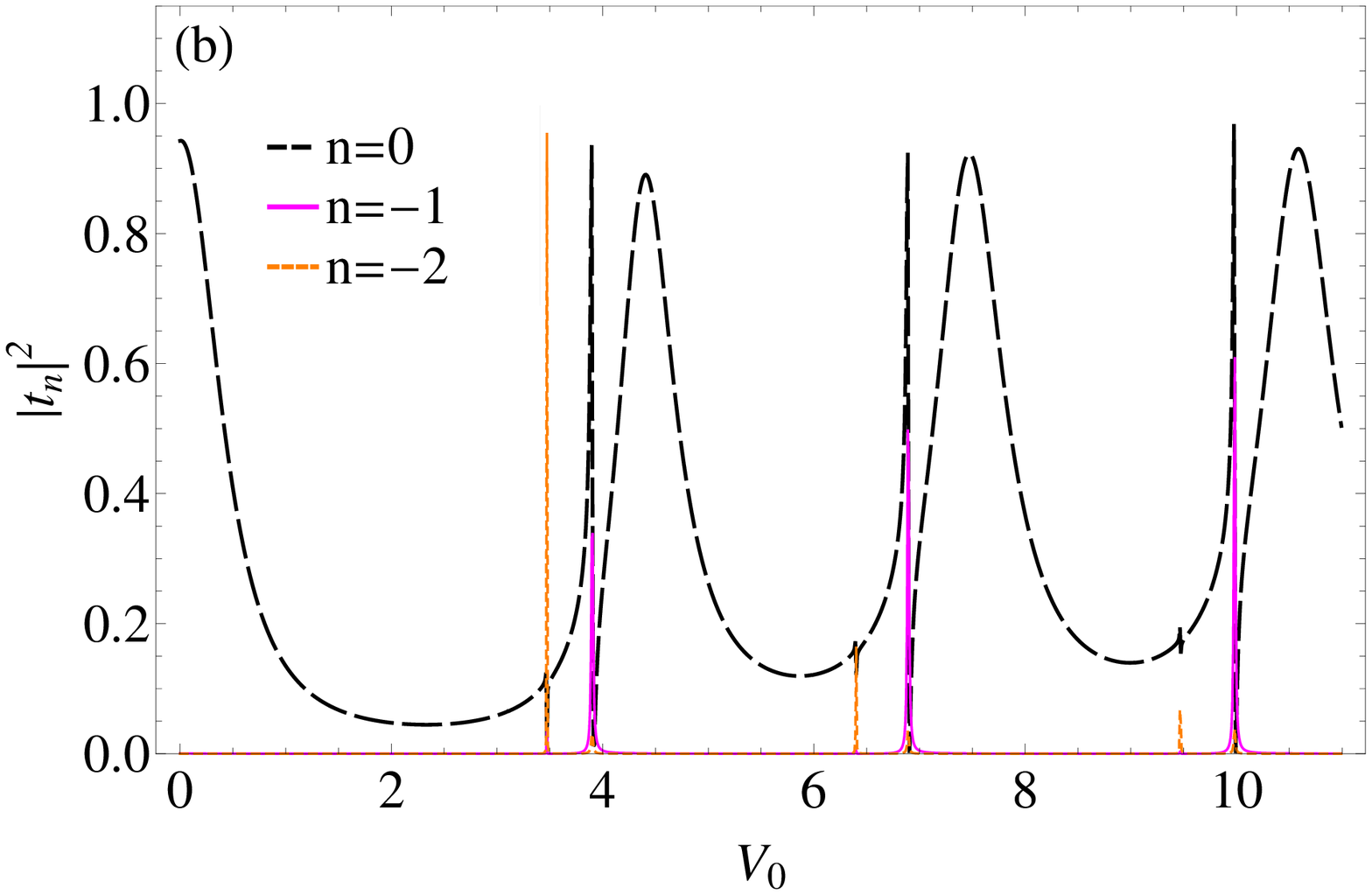}
\caption{Panel (a): $\left\langle T \right\rangle$ as a function of barrier height $V_0$ for the indicated values of the oscillation amplitude $\Omega$.
Additional parameters: $E_0=1.1 mc^2, \nu=0.2, L=1$.
Panel (b): Sideband transmission amplitudes $t_n$ as a function of barrier height $V_0$, for comparison with the $\Omega=0.2$ curve of panel (a). Note that the curves corresponding to $n=-1$ and $n=-2$ have been rescaled (divided by a factor of $160$).}
\label{fanofig1}
\end{figure}

\begin{figure}[tbh]
\includegraphics[width=8.0cm]{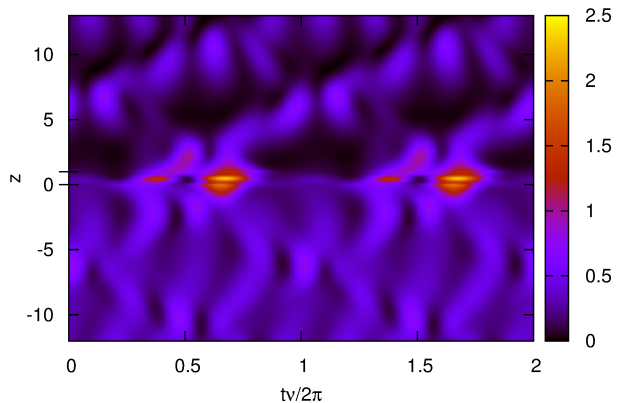}
\caption{Density plot of $\rho(z,t)$, for $L=1$, $V_0=3.97 mc^2$, $E_0=1.1 mc^2$ and $\Omega=1.4$.}
\label{fanofig2}
\end{figure}
In the following we focus on positive values of $V_0,$ as it is depicted on Fig.~\ref{firstfig}. For $L=1,$ one usually finds a single bound state of the static potential barrier, which simplifies the interpretation of the results. The energy eigenvalues $\epsilon$ corresponding to these states are in the range of $\left[-mc^2, mc^2\right],$ thus they are evanescent in regions 1 and 3, where the potential is zero. Inspecting the positions of the resonances in Fig.~\ref{fanofig1}, we find that they appear when
\begin{equation}
E_0+n \hbar\nu\approx\epsilon,
\label{resonance}
\end{equation}
with $n$ being integer. Note that we cannot expect exact equality in Eq.~(\ref{resonance}), since AC Stark shift modifies the energy levels. In the investigated parameter ranges, the relative difference of the left end right hand sides of Eq.~(\ref{resonance}) was around $5\%$ for low barriers (in the sense that $V_0$ is not considerably larger than $\hbar\,\Omega$), and it decreased for larger values of $V_0.$

As one can expect, resonances corresponding to increasing magnitudes of $n$ are weaker. According to the terminology that is often used -- in spite of the fact that the potential oscillations are not quantized -- we may say that the probability of higher order processes that involve 2, 3, etc.~excitation quanta ("photons") are considerably lower than that of "single photon" processes. However, when the amplitude of the oscillation increases, higher order resonances get more pronounced.

Panel b) of Fig.~\ref{fanofig1} shows a clear example that whenever there is a resonance in $\langle T\rangle,$ the transmitted amplitude $t_0$ corresponding to the incoming electron's energy has a sharp minimum. At the same value of $V_0,$ one of the coefficients $t_n$ (that belong to $\omega_n=E_0/\hbar+n\nu,$ with $n<0$) has a peak. We can clearly identify the peaks corresponding to $n=-1$ and $-2$ in Fig.~\ref{fanofig1} b).

As an example for a third order resonance, Fig.~\ref{fanofig2} shows $\rho(z,t)$ for parameter values where $t_{-3}$ has the highest magnitude amongst all transmission coefficients. As we can see, the fact that the incoming electron excites a localized bound state appears in this figure as an increase of the electron density. Note that this temporal trapping in region 2 is considerably more pronounced than in the cases seen in Figs.~\ref{fig6} and \ref{fig7}; the maximal value of $\rho(z,t)$ inside the oscillating potential is five time larger than anywhere else.

Figs.~\ref{fanofig1} and \ref{fanofig2} demonstrate that the abrupt changes in the cycle averaged transmission and reflection probabilities
are due to Fano-type resonances as described by Eq.~(\ref{resonance}). Note that this kind of behavior can appear in principle for higher energies as well, but then the order of the process [the values of $n$ in Eq.~(\ref{resonance})] is so high, that it means only a practically invisible correction for the transmission probabilities.

Finally, let us analyze to what extent the effects presented so far are relevant for graphene. As we have mentioned earlier, according to Ref.~[\onlinecite{nat_mat_2003}] a band gap $\Delta$ as large as 0.26\,eV can appear in epitaxially grown graphene on SiC substrate. This band gap plays the role of the energy difference of $2mc^2$ that separates positive and negative energy eigenstates of the massive Dirac equation we considered. Our findings are relevant when $\hbar\,\Omega,$ $V_0$ and $\hbar\nu$ has the same order of magnitude as $\Delta.$ The characteristic frequency $\nu/2\pi$ is in the THz regime ($\hbar\nu\approx\Delta/2$ gives $\nu=25$ THz), which is in the experimentally achievable range. That is, generation of wave packets, the existence of alternating relativistic currents in the reflected region as well as the appearance of Fano-type resonances can be visible also in the case of the graphene.

\section{Summary}
We presented a theoretical work on the relativistic scattering of massive Dirac particles on a time-periodic rectangular potential barrier, using Floquet theory. We used the dependence of the cycle-averaged transmission probability on the barrier height to describe the quasistationary behavior of the system, in the case of a weakly relativistic, relativistic and ultra-relativistic incident particle. In each case, the system shows Klein tunneling and Fano-type resonances. We explored the details of the transport with the help of space and time dependent currents and densities, which show explicitly that the oscillating barrier generates wavepackets from the incident plane wave. We explained in detail the Fano-type resonances with the interplay of the sideband states generated by the oscillating potential and the bound states of the barrier, and showed the corresponding temporal population trapping in the barrier region. Finally we discussed the relevance of our results to graphene with an induced bandgap.

\section{Acknowledgments}

This work was partially supported
by the European Union and the European Social Fund through projects
''Supercomputer, the national virtual lab'' (grant no.: TAMOP-4.2.2.C-11/1/KONV-2012-0010)
and
''Impulse lasers for use in materials science and biophotonics'' (grant no.: TAMOP-4.2.2.A-11/1/KONV-2012-0060),
and by the Hungarian Scientific Research Fund (OTKA) under Contract No.~T81364.

\end{document}